\documentclass[aps,twocolumn,pra,superscriptaddress,showpacs,tightenlines]{revtex4}
\usepackage{amssymb}
\usepackage{amsmath}
\usepackage{graphicx}
\usepackage{epsfig}
\usepackage{subfigure}
\usepackage{amsfonts}

\begin{document}

\title{Enhanced Deflection of Light Ray by Atomic Ensemble on Coherent
Population Oscillation}
\author{Jing Lu}
\affiliation{Key Laboratory of Low-Dimensional Quantum Structures and Quantum Control of
Ministry of Education, and Department of Physics, Hunan Normal University,
Changsha 410081, China}
\author{Lan \surname{Zhou}}
\affiliation{Key Laboratory of Low-Dimensional Quantum Structures and Quantum Control of
Ministry of Education, and Department of Physics, Hunan Normal University,
Changsha 410081, China}
\author{Le-Man Kuang}
\affiliation{Key Laboratory of Low-Dimensional Quantum Structures and Quantum Control of
Ministry of Education, and Department of Physics, Hunan Normal University,
Changsha 410081, China}

\begin{abstract}
In recent experiments[e.g., Nature Physics 2, 332 (2006)], the enhanced
light deflection in an atomic ensemble due to inhomogeneous fields is
demonstrated by the electromagnetically induced transparency (EIT) based
mechanism. In this paper, we explore an different mechanism for the similar
phenomenon of the enhanced light deflection. This mechanism is based on the
coherent population oscillation, which leads to the hole burning in the
absorption spectrum. The medium causing the deflection of probe light is an
ensemble of two-level atoms manipulated by a strong controlled field on the
two photon resonances. In the large detuning condition, the response of the
medium to the pump field and signal field is obtained with steady state
approximation. And it is found that after the probe field travels across the
medium, the signal ray bends due to the spatial-dependent profile of the
control beam.
\end{abstract}

\maketitle

\section{\label{sec:one}Introduction}

Recently, slow light propagation has attracted a great deal of
attention due to the fundamental aspects of nontrivial light pulse
manipulation~\cite{Harris-1} and possible applications for optical
delay lines~\cite{deline-1} , for quantum computing and quantum
communication \cite{Lukin-1}, and for developing sensitive
measurement techniques~\cite{fleisch-1}. The experiments for slow
light are carried out on various types of materials such as cold
sodium atoms~\cite{caslow-1,caslow-2}, atom vapors~\cite
{Lukin-3,Lukin-4,harris-2,harris-3}, and solids state system\cite
{solid-1,solid-2,solid-3,solid-4,solid-5}.

The physical mechanism of slow light is to create a narrow
transparency window (burn a hole) within the absorption line by an
intense coupling laser field. Also a sharp normal dispersion of the
refractive index is accompanied within this narrow window. Thus both
a low group velocity of light and an enhanced transmission of light
can be achieved. A basic way to produce slow light is
electromagnetically induced transparency (EIT)~\cite
{Harris-1,Lukin-1,caslow-1,harris-4}. The EIT effect usually happens
in the so-called $\Lambda $-type atomic system, which contains two
lower states with separate couplings to an excited state through two
electromagnetic fields (probe and control light). When the
absorption of light by both transitions is suppressed due to
destructive interference between excitation pathways to the upper
level, the medium becomes transparent with respect to the probe
field. The group velocity of the light depends on the parameters of
the control field~\cite{sprl91,spra69}.

Another method to producing slow light is the coherent population
oscillation (CPO)~\cite{deline-1,solid-2,solid-4,solid-5,pco-1}. In
this sense, the transition is excited by a probe field and a control
light from the ground state to the excited state of a two-level
system. The beating due to the slight detuning between pump and
probe lights leads to a periodic modulation of the atomic
population, which create a narrow spectral dip in the probe
absorptive spectrum. Due to its insensitive to the dephasing of the
atomic coherence, the slow light has been found at room
temperature~\cite{deline-1}, and its speed can be reduced as low as
a few tens of meters per second~\cite{solid-2}.

Actually, the conventional studies of slow-light phenomenon focus on
various effects in the time-domain, but most recently, much
attention has been paid to light propagation in the spatial domain.
The effect of an external field with spatially inhomogeneous
distribution has been studied~\cite
{lightd-1,lightd-2,lightd-3,lightd-4}. It is found that the light
ray bends when a magnetic field with small gradient vertical to the
propagation direction is applied to a EIT medium~\cite{lightd-1}.
Also in a EIT medium, the light deflection is explicitly observed
when the atomic ensemble is driven by a optical field with an
inhomogeneous profile along transverse direction~\cite{lightd-2}.

For the slow light phenomena due to CPO, almost all theoretical treatments
are confined to the homogeneous transverse spatial distribution of the
control field. In this paper, we address how the spatial profile of the
control field along transverse direction affects the propagation of the
probe beam for the CPO based slow light . Here we treat the external fields
classically. After investigating the atomic response to the control field
and the probe field by the perturbation approach, the spatial motions of the
control field is governed by an effective nonlinear Schr\"{o}dinger equation
in the low intensity limit. It is found that the control field can propagate
for long distances with an invariant transverse profile. And with an
effective potential induced by the steady atomic response, the spatial
motions of the probe field is described by an effective Schr\"{o}dinger
equation in the limits of linear response. And therefore the deflection of
the light ray comes out straightforwardly.

This paper is organized as follow: in sec.~\ref{sec:two}, we present the
theoretical model for a two-level atomic ensemble interacting with a control
field and a much weaker probe field. In Sec.~\ref{sec:three}, the
perturbation theory is applied to obtain the atomic motion equation which is
related to the response to the external field. In Sec.~\ref{sec:four}, we
derive a system of equations which govern the spatial motion of the control
field and the probe field in the optically controlled atomic medium. Then
the deflection of the probe beam is investigated by the transverse spatial
profile of the control light. In Sec.~\ref{sec:sum}, we make our conclusion.

\section{\label{sec:two}Optically controlled two level atoms}

\begin{figure}[tbp]
\includegraphics[width=4 cm]{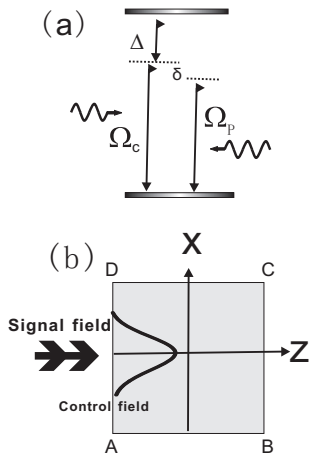}
\caption{(a) Atomic two-level system interacting with the probe light and a
control light. $\Omega _{i}$ denotes the Rabi frequency of the control and
probe field. (b) Configuration of the optical beams inside the atomic medium.
}
\label{fig:1}
\end{figure}

The system for producing slow light by CPO involves three subsystems: the
atomic medium, a control field and a weaker signal field. The atomic medium
is an ensemble of $N$ identical atoms confined in a cell ABCD shown in Fig.~%
\ref{fig:1}(b).
Each of atoms has two internal states -- the ground state
$|g\rangle$ and the excited state $|e\rangle $. The atomic
transition is driven by two optical fields, a weaker probe field and
a stronger control field, as shown in the top panel of
Fig.~\ref{fig:1}. The energy difference between $ \left\vert
g\right\rangle $ and $\left\vert e\right\rangle $ is denoted by $
\omega _{eg}=\omega _{e}-\omega _{g}$. $\Delta =\nu _{c}-\omega
_{eg}$ is the detuning between the atomic transition and the control
field with carried frequency $\nu _{c}$. In this literature, large
detuning is considered. The probe field is slightly detuned from the
control field, and it is denoted by $\delta =\nu _{c}-\nu _{p}$. The
detuning between the control and probe light leads to periodic
modulation of the two-level system population oscillation, which
substantially reduces the group velocity~\cite{solid-4,solid-5}.

As shown in Fig.~\ref{fig:1}(b), both the probe field and control field
propagate parallel along the $z$-direction with wave number $k$ and $k_{c}$
respectively. The Hamiltonian
\begin{equation}
H=H^{(A)}+H^{(I)}\text{.}  \label{2-01}
\end{equation}
is composed of two terms. To describe the atomic medium, we introduce the
collective atomic operator~\cite{Lukin-3}
\begin{equation}
\tilde{\sigma}_{\mu \nu }\left( r,t\right) =\frac{1}{N_{r}}\sum_{r_{j}\in N}%
\tilde{\sigma}_{\mu \nu }^{j}\left( t\right) \text{,}  \label{2-01a}
\end{equation}%
which average over a small but macroscopic volume $V$ containing many atoms $%
N_{r}=\left( N/V\right) dr\gg 1$ around position $r$, where $N$ is the total
number of atoms. Then the unperturbed atomic Hamiltonian reads
\begin{equation}
H^{(A)}=\frac{N}{V}\int d^{3}r\left( \omega _{e}\tilde{\sigma}_{ee}+\omega
_{g}\tilde{\sigma}_{gg}\right),  \label{2-02}
\end{equation}%
where we have neglected the kinetic term of atoms. The interaction
Hamiltonian $H^{(I)}$ describes the coupling between the external field and
the atomic ensemble. Under the electric-dipole approximation and the
rotating-wave approximation, the interaction Hamiltonian $H^{(I)}$ reads
\begin{equation}
H^{(I)}=\frac{N}{V}\int d^{3}r\left( \tilde{\Omega}_{p}\tilde{\sigma}_{eg}+%
\tilde{\Omega}_{c}\tilde{\sigma}_{eg}+h.c.\right)  \label{2-03}
\end{equation}
where
\begin{equation}
\tilde{\Omega}_{j}=d_{eg}\tilde{E}_{j}^{+}/\hbar \text{, \ }j=p,c
\label{2-04}
\end{equation}%
are the Rabi frequencies associated with the corresponding external
fields. $\tilde{E}_{j}^{+}$ is the positive frequency part of the
corresponding external field.

By introducing the slow varying variables for both fields
\begin{subequations}
\label{2-05}
\begin{eqnarray}
\tilde{E}_{p}^{+} &=&E_{p}\left( r,t\right) e^{i\left( k_{p}z-\nu
_{p}t\right) }\text{,} \\
\tilde{E}_{c}^{+} &=&E_{c}\left( r\right) e^{i\left( k_{c}z-\nu _{c}t\right)
}\text{,}
\end{eqnarray}
and for atomic transition operator
\end{subequations}
\begin{equation}
\tilde{\sigma}_{eg}=\sigma _{eg}e^{-ik_{c}z}  \label{2-06}
\end{equation}%
the dynamics of this system is described by the interaction Hamiltonian
\begin{eqnarray}
H_{I} &=&\frac{N}{V}\int d^{3}r[\frac{d_{eg}}{\hbar }\left( E_{p}e^{i\left(
\delta t-k_{\delta }z\right) }+E_{c}\right) \sigma _{eg}  \label{2-07} \\
&&+h.c.-\Delta \sigma _{ee}]  \notag
\end{eqnarray}
in the rotating reference frame, where $k_{\delta }=k_{c}-k_{p}$ and $d_{eg}$
is dipole transition element. Here we have assumed that the control beam is
monochromatic wave, however the slow varying variable $E_{p}$ for the probe
light is a superposition of monochromatic waves, which have a small
frequency variance $\Delta\nu$ around the mean frequency $\nu _{p}$.

\section{\label{sec:three}atomic response}

It is well-known that, when atoms are subjected to an electric field, the
applied field displaces the positive charges and the negative charges in
atoms from their usual positions. This small movement that positive charges
in one direction and negative ones in the other will result in collective
induced electric-dipole moments. Every dipole behaves collectively to give a
response to the light. We now investigate the response of the two level
atomic ensemble driven by a monochromatic classical field. The equations for
the atomic coherence are described by the density matrix elements
\begin{subequations}
\label{3-01}
\begin{eqnarray}
\dot{\sigma}_{ee} &=&i\frac{d_{eg}^{\ast }}{\hbar }V_{eg}\sigma _{ge}+h.c. \\
\dot{\sigma}_{gg} &=&i\frac{d_{eg}}{\hbar }V_{eg}^{+}\sigma _{eg}+h.c. \\
\dot{\sigma}_{eg} &=&-i\Delta \sigma _{eg}-i\frac{d_{eg}^{\ast }}{\hbar }%
V_{eg}\left( \sigma _{ee}-\sigma _{gg}\right) \text{.}
\end{eqnarray}%
where
\end{subequations}
\begin{equation}
V_{eg}=E_{p}^{+}e^{i\left( k_{\delta }z-\delta t\right) }+E_{c}^{+}
\label{3-02}
\end{equation}%
is the amplitude of the applied optical field. We consider the ensemble of
closed two-level atoms. Due to the spontaneous emission, the population of
the upper level will decay. We assume that the upper level $\left\vert
e\right\rangle $ decays to the lower level $\left\vert g\right\rangle $ at a
rate $\gamma _{1}$, which means the lifetime of the upper level is given by $%
T_{1}=\gamma _{1}^{-1}$. As the two-level is closed, any population that
leaves the upper level enters the ground level. Thus $T_{1}$ is the life
time of the population difference of the ground $\left\vert g\right\rangle $
and excited $\left\vert e\right\rangle $ states. We also assume that the
atomic dipole moment dephases at a rate $\gamma _{2}=T_{2}^{-1}$. By adding
decay terms phenomenologically, the dynamic of this atomic ensemble is
described by
\begin{subequations}
\label{3-03}
\begin{eqnarray}
\dot{w} &=&-\gamma _{1}\left( w-w_{eq}\right) +i\frac{2d_{eg}^{\ast }}{\hbar
}V_{eg}\sigma _{ge}+h.c. \\
\dot{\sigma}_{eg} &=&-\left( i\Delta +\gamma _{2}\right) \sigma _{eg}-i\frac{%
d_{eg}^{\ast }}{\hbar }V_{eg}w\text{.}
\end{eqnarray}%
where $w$ is the population inversion operator defined as
\end{subequations}
\begin{equation}
w=\sigma _{ee}-\sigma _{gg}\text{;}  \label{3-04}
\end{equation}%
$w_{eq}$ is the population difference in thermal equilibrium.

Eq.~(\ref{3-03}) cannot readily be solved exactly. As the probe
field is much weaker than the control field, we seek a solution to
the equations of the atomic motion that is correct to all orders in
the amplitude of the control field and is correct to lowest order in
the amplitude of the probe field. Though the polarization induced by
the external fields in the ensemble of two-level atoms is described
by the equation

\begin{equation}
P\left( r,t\right) =\frac{N}{V}\sum_{j}d_{eg}^{\ast }\left\langle \sigma
_{ge}\right\rangle e^{i\left( k_{j}z-\nu _{j}t\right) }\text{,}  \label{3-05}
\end{equation}%
which generally includes components not only at the field frequencies $\nu
_{c}$ and $\nu _{p}$ but also at $\nu _{p}\pm k\left( \nu _{c}-\nu
_{p}\right) $ with $k$ integral, only the frequencies $\nu _{c}$, $\nu _{p}$
and $\nu _{c}+\delta $ will occur in the atomic response to lowest order of
the amplitude $E_{p}$. Thus we treat the atomic equations perturbatively,
and require that the steady-state solution of Eq.~(\ref{3-03}) be of the
form~\cite{solid-4,pco-1}
\begin{subequations}
\label{3-06}
\begin{eqnarray}
\sigma _{eg} &=&\sigma _{eg}^{(0)}+\sigma _{eg}^{(+)}e^{i\left( k_{\delta
}z-\delta t\right) }+\sigma _{eg}^{(-)}e^{-i\left( k_{\delta }z-\delta
t\right) }\text{,} \\
w &=&w^{(0)}+w^{(+)}e^{i\left( k_{\delta }z-\delta t\right)
}+w^{(-)}e^{-i\left( k_{\delta }z-\delta t\right) }\text{,}
\end{eqnarray}%
where $w^{(0)}$ and $\sigma _{eg}^{(0)}$ are the population and
polarization in the absence of the probe field, and other terms are
in the first order of $E_{p}$. Actually, Eqs.~(\ref{3-06}) is a
truncation of the Floquet basis expansion~\cite{Floquet}. The
dynamics of population and dipole moment of the atomic ensemble are
given by the equations in the zeroth order of $E_{p}$
\end{subequations}
\begin{subequations}
\label{3-07}
\begin{eqnarray}
\dot{w}^{(0)} &=&-\gamma _{1}w^{(0)}+\gamma _{1}w_{eq}+2i\left( \Omega
_{c}^{\ast }\sigma _{ge}^{(0)}-h.c.\right) \\
\dot{\sigma}_{eg}^{(0)} &=&-\left( i\Delta +\gamma _{2}\right) \sigma
_{eg}^{(0)}-i\Omega _{c}^{\ast }w^{(0)} \\
\dot{\sigma}_{ge}^{(0)} &=&\left( i\Delta -\gamma _{2}\right) \sigma
_{ge}^{(0)}+i\Omega _{c}w^{(0)}
\end{eqnarray}%
and the first order of $E_{p}$
\end{subequations}
\begin{subequations}
\label{3-08}
\begin{eqnarray}
\dot{w}^{(-)} &=&\left( i\delta -\gamma _{1}\right) w^{(-)}-i2\Omega
_{p}\sigma _{eg}^{(0)} \\
&&+i2\left( \Omega _{c}^{\ast }\sigma _{ge}^{(-)}-h.c.\right)  \notag \\
\dot{\sigma}_{eg}^{(-)} &=&\left[ i\left( \delta -\Delta \right) -\gamma _{2}%
\right] \sigma _{eg}^{(-)}-i\Omega _{c}^{\ast }w^{(-)} \\
\dot{\sigma}_{ge}^{(-)} &=&\left[ i\left( \delta +\Delta \right) -\gamma _{2}%
\right] \sigma _{ge}^{(-)} \\
&&+i\Omega _{p}w^{(0)}+i\Omega _{c}w^{(-)}  \notag
\end{eqnarray}%
where we have introduce the Rabi frequencies
\end{subequations}
\begin{equation}
\Omega _{j}=\frac{d_{eg}}{\hbar }E_{j}\text{, }j=c,p.  \label{3-09}
\end{equation}
Under the adiabatic approximation that the evolution of the atomic
system is much faster than the temporal change of the radiation
field, we set the left hand sides of Eqs.~(\ref{3-07})
and~(\ref{3-08}) to zero. Therefore, the population and dipole
moment in the zeroth order of $E_{p}$ read
\begin{subequations}
\label{3-10}
\begin{eqnarray}
w^{(0)} &=&\frac{\gamma _{1}\left( \Delta ^{2}+\gamma _{2}^{2}\right) w_{eq}%
}{\gamma _{1}\left( \Delta ^{2}+\gamma _{2}^{2}\right) +4\gamma
_{2}\left\vert \Omega _{c}\right\vert ^{2}} \\
\sigma _{ge}^{(0)} &=&\frac{\gamma _{1}\left( i\gamma _{2}-\Delta \right)
\Omega _{c}}{\gamma _{1}\left( \Delta ^{2}+\gamma _{2}^{2}\right) +4\gamma
_{2}\left\vert \Omega _{c}\right\vert ^{2}}w_{eq}
\end{eqnarray}%
and the dipole moment in the first order of the probe field amplitude reads
\end{subequations}
\begin{eqnarray}
\sigma _{ge}^{(-)} &=&-\frac{D\left( i\gamma _{2}-\Delta \right) }{D\left(
i\gamma _{2}-\Delta \right) \left( \delta +\Delta +i\gamma _{2}\right) }%
w^{(0)}\Omega _{p}  \label{3-11} \\
&&-\frac{2\left\vert \Omega _{c}\right\vert ^{2}\left( \delta +2i\gamma
_{2}\right) \left( \delta -\Delta +i\gamma _{2}\right) }{D\left( i\gamma
_{2}-\Delta \right) \left( \delta +\Delta +i\gamma _{2}\right) }%
w^{(0)}\Omega _{p}  \notag
\end{eqnarray}%
where
\begin{eqnarray}
D &=&\left( \delta +i\gamma _{1}\right) \left( \delta -\Delta +i\gamma
_{2}\right) \left( \delta +\Delta +i\gamma _{2}\right)  \label{3-12} \\
&&-4\left\vert \Omega _{c}\right\vert ^{2}\left( i\gamma _{2}+\delta \right)
\notag
\end{eqnarray}%
Here $\sigma _{ge}^{(0)}$ and $\sigma _{ge}^{(-)}$ determine the atomic
response to the control and probe fields respectively.

\section{\label{sec:four}Propagation of lights in an two-level atomic medium}

We now consider the deflection of the probe light in the rectangular
medium due to the transverse spatial profile of the control light
shown in Fig.~(\ref{fig:1}b). For a nonmagnetic medium with no free
charges and no free currents, a driven wave equation
\begin{equation}
\nabla ^{2}\tilde{E}-\varepsilon _{0}\mu _{0}\frac{\partial ^{2}}{\partial
t^{2}}\tilde{E}=\mu _{0}\frac{\partial ^{2}}{\partial t^{2}}P  \label{4-01}
\end{equation}%
can be obtained from the Maxwell equations, where $\nabla ^{2}$ is the
Laplacian operator, and $\tilde{E}$ is the sum of amplitudes of the control
and probe beam. Here the dielectric response $P$ acts as an effective source
to generate the electromagnetic field. In the paraxial approximation and
slowing varying amplitude approximation, the propagating equations for the
external classical field amplitude $\tilde{E}^{+}=\tilde{E}_{c}^{+}+\tilde{E}%
_{p}^{+}$ read
\begin{subequations}
\label{4-02}
\begin{eqnarray}
i\partial _{z}\Omega _{c}+\frac{1}{2k_{c}}\nabla _{T}^{2}\Omega _{c} &=&-%
\frac{N}{c}\left\vert g_{c}\right\vert ^{2}\sigma _{ge}^{(0)} \\
i\partial _{t}\Omega _{p}+ic\partial _{z}\Omega _{p}+\frac{c}{2k_{p}}\nabla
_{T}^{2}\Omega _{p} &=&-N\left\vert g_{p}\right\vert ^{2}\sigma _{ge}^{(-)}
\end{eqnarray}%
where $\nabla _{T}^{2}$ is the transverse laplacian and
\end{subequations}
\begin{equation}
g_{j}=d_{eg}\sqrt{\frac{v_{j}}{2\varepsilon _{0}\hbar V}}\text{, }j=c,p
\label{4-03}
\end{equation}%
The right hands of Eqs.~(\ref{4-02}) are the response functions of the atoms
located in position $r$.

For simplification, we deal with our problem in two-dimension, that
is in the x-z plane. Such a system can be realized experimentally by
putting the atomic ensemble in a planar waveguide. In the large
detuning between the control light and the atomic transition, most
atoms are in the ground state $\left\vert g\right\rangle $, which
means the population difference in thermal equilibrium $w_{eq}=-1$.
As the detuning usually is much larger than the dephasing rate
$\gamma _{2}$, we can neglect the imaginary part of the atomic
response to the optical fields, then the propagating equations read
\begin{subequations}
\label{4-04}
\begin{eqnarray}
i\partial _{z}\Omega _{c}+\frac{1}{2k_{c}}\partial _{x}^{2}\Omega _{c} &=&%
\frac{\alpha _{c}\Omega _{c}}{1+2\beta \left\vert \Omega _{c}\right\vert ^{2}%
}\text{,} \\
i\partial _{t}\Omega _{p}+ic\partial _{z}\Omega _{p}+\frac{c}{2k_{p}}%
\partial _{x}^{2}\Omega _{p} &=&\frac{\alpha _{p}\Omega _{p}}{\left(
1+2\beta \left\vert \Omega _{c}\right\vert ^{2}\right) ^{2}}\text{.}
\end{eqnarray}
where $\delta =0$ has been set and
\end{subequations}
\begin{subequations}
\label{4-05}
\begin{eqnarray}
\alpha _{c} &=&-\frac{N}{c}\left\vert g_{c}\right\vert ^{2}\frac{\Delta }{%
\Delta ^{2}+\gamma _{2}^{2}}\text{,} \\
\alpha _{p} &=&-N\left\vert g_{p}\right\vert ^{2}\frac{\Delta }{\left(
\gamma _{2}^{2}+\Delta ^{2}\right) }\text{,} \\
\beta &=&\frac{2\gamma _{2}}{\gamma _{1}\left( \gamma _{2}^{2}+\Delta
^{2}\right) }\text{.}
\end{eqnarray}%
The consideration for taking $\delta =0$ reflects the principle of
this problem by the following reasons: 1) a resonance occurs at
$\delta =0$ which can be obtained by the inspection of
Eq.~(\ref{3-12}); 2) Though two-level atoms oscillate at frequency
$\delta$, it becomes important only when $\delta $ is less than
$\gamma _{1}$, which results in the occurrence of a narrow spectral
hole with width proportional to $\gamma _{1}$ burning in the
homogeneous absorption profile of the probe field. Hence the group
velocity of the probe beam is somewhat reduced by CPO.

As the atomic response to the probe beam depends on the control beam, we
first deal with the propagating equation for the control beam. In the limit
of low intensities of the control beam, the control wave experiences a
refractive index correlated with the optical intensity
\end{subequations}
\begin{equation}
i\partial _{z}\Omega _{c}+\frac{1}{2k_{c}}\partial _{x}^{2}\Omega
_{c}=\alpha _{c}\left( 1-2\beta \left\vert \Omega _{c}\right\vert
^{2}\right) \Omega _{c}\text{,}  \label{4-06}
\end{equation}%
i.e. the control wave induces a purely third-order nonlinear optical
response. By defining the momentum operators $p_{\mu }=-i\partial
_{\mu }$ and the effective mass $m_{p}=k$, Eq.~(\ref{4-06}) is a
nonlinear Schr\"{o}dinger-like equation, which has a soliton
solution
\begin{equation}
\Omega _{c}=\frac{\sqrt{Q}}{2}e^{i\left( \frac{Q^{2}}{4}-\alpha _{c}\right)
z}\text{sech}\left( \frac{x}{L_{c}}\right)  \label{4-07}
\end{equation}%
with $Q=\alpha _{c}\beta $. Such solitary wave can propagate for long
distances with an invariant transverse profile. Here
\begin{equation}
L_{c}=\frac{\sqrt{2/k_{c}}}{\alpha _{c}\beta }  \label{4-08}
\end{equation}%
is the transverse size of the control beam.

We now consider the propagation of an initial Gaussian probe wavepacket
\begin{equation}
\Omega _{p}\left( 0,x,z\right) =\frac{1}{\sqrt{\pi b^{2}}}\exp \left[ -\frac{%
\left( x-a\right) ^{2}+z^{2}}{b^{2}}\right]   \label{4-09}
\end{equation}%
in this two-level medium, where $b$ ($<L_{c}$) is the width of the
probe field and $a$ is the initial wave packet center of the probe
field along $x$-direction. The magnitude $\left\vert a\right\vert $
denotes the distance from the center of the control beam. In order
to investigate the evolution of this initial state, we expand
$\left( 1+2\beta \left\vert \Omega _{c}\right\vert ^{2}\right)
^{-2}$ at the position $a$ and retain the linear term proportional
to $x-a$. Then the paraxial equation for the probe beam becomes
\begin{equation}
i\partial _{t}\Omega _{p}+ic\partial _{z}\Omega _{p}+\frac{c}{2k_{p}}%
\partial _{x}^{2}\Omega _{p}=\left[ \eta _{0}+\eta _{1}\left( x-a\right) %
\right] \Omega _{p}  \label{4-10}
\end{equation}%
where parameters
\begin{subequations}
\label{4-11}
\begin{eqnarray}
\eta _{0} &=&\alpha _{p}\left[ 1+\frac{\beta Q}{2}\text{sech}^{2}\left(
\frac{a}{L_{c}}\right) \right] ^{-2} \\
\eta _{1} &=&\alpha _{p}Q^{2}\beta \sqrt{2k_{c}}\frac{\text{sech}^{2}\left(
\frac{a}{L_{c}}\right) \tanh \left( \frac{a}{L_{c}}\right) }{\left[ 1+\frac{%
\beta Q}{2}\text{sech}^{2}\left( \frac{a}{L_{c}}\right) \right] ^{3}}
\end{eqnarray}%
Also by defining the momentum operators $p_{\mu }=-i\partial _{\mu }$ and
the effective mass $m_{p}=k_{p}/c$, the dynamic of the probe wave $\Omega
_{p}$ is governed by the Schr\"{o}dinger-like equation $i\partial _{t}\Omega
_{p}=H\Omega _{p}$ with the effective Hamiltonian
\end{subequations}
\begin{equation}
H=cp_{z}+\frac{1}{2m_{p}}p_{x}^{2}+\eta _{0}+\eta _{1}\left( x-a\right)
\text{.}  \label{4-12}
\end{equation}%
Thus by making use of the Wei-Norman algebraic
method~\cite{lightd-3,swna} (see the appendix), it can be found
that, after passing through the Rb gas cell, the initial center
$\left( x,z\right) =\left( a,0\right) $ of the probe field is
shifted to
\begin{subequations}
\label{4-13}
\begin{eqnarray}
x &=&a-\frac{\eta _{1}L^{2}}{2kc} \\
z &=&L
\end{eqnarray}%
By tracking the center motion of the probe beam, a mirage effect
might occur to the probe light, which is determined by the sign of
the detuning $\Delta $ and the initial center $a$ of the probe beam.
When the center is collinear to that of the control field
$a=x_{0}=0$, the trajectory of the signal light is a straight line.
We assign the positive sign for $a$ as the probe beam shifted to the
right with respect to the center of control light, and denote $a<0$
as the signal beam shifted to the left. When the probe beam is
shifted to the right, i.e. $a>0$, for red detuning $\Delta <0$, the
probe wave feels a ``attractive potential'' within the atomic medium
due to $\eta _{1}>0$, thus the trajectory bends into the left side
of the $\hat{e}_{z}$-axis; for blue detuning $\Delta >0$, the signal
light undergo an ``repulsion potential'' due to $\eta _{1}<0$,
therefore the ray bends to the right. When the probe beam is shifted
to the left, i.e. $a<0$, as the coefficient of the linear potential
is larger than zero, that is $\Delta >0$, the probe beam experiences
a ``attractive force'' within the two-level atomic medium, and its
center is shifted to the left. As $\eta _{1}$ is smaller than zero,
i.e. $\Delta <0$, the probe beam suffers a ``repulsion force''
during passing through the atomic medium, hence the whole wave
packet goes to the right. The corresponding schematic diagram is
given in Fig.\ref{fig:2}, where dark thick line is the transverse
spatial profile of the control light.
\begin{figure}[tbp]
\includegraphics[width=6 cm]{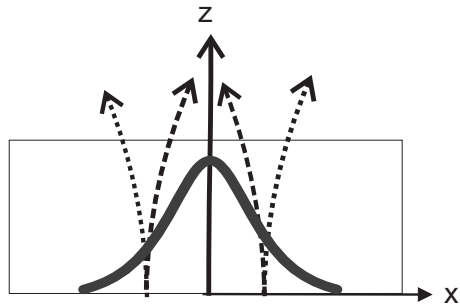}
\caption{ Schematic illustration about the ray deflection of the
probe light in the presence of spatial-distributed coupling light.
The dark thick line is the transverse spatial profile of the control
light.} \label{fig:2}
\end{figure}
The dash lines give the deflection at $\Delta <0$, the dotted lines describe
the light trajectory at $\Delta >0$ and the black solid depict the light ray
at $a=0$.

\section{\label{sec:sum}Conclusions}

In conclusion, we have theoretically predicted a phenomenon of the enhanced
light deflection by an atomic ensemble through coherent population
oscillation mechanism, which is realized by a two-level atomic ensemble
interacting with a control field and a much weaker probe field. After
calculating the atomic response to the external fields by the perturbation
approach, we obtain a system of equations that describe the spatial motion
of the control field and the probe field in the two-level atomic medium. The
propagation of the control field is governed by a nonlinear Schrodinger
equation in the limit of a low intensity. It shows that a solitary wave can
be excited, which has an invariant transverse profile for a long-distance
propagation. Due to a transparency window with a width of the order of $%
\gamma _{1}$ and the achievement of the substantially reduced group velocity
for the probe field by CPO, we found that the deflection of the light ray
can be controlled by two controllable external parameters: the initial
center of the probe beam with respect to the control light, and the detuning
between the control field and the atomic transition, which is similar to
that in EIT medium investigated by Ref.~\cite{lightd-3,lightd-4}. Our
analysis maybe provide a new technique for transverse light guiding and it
is much appreciated that our prediction can be verified in the further
experiment.

This work was supported by the NSFC with Grant No 10775048, No.
10704023, No. 10775048, and No. 10325523, and NFRPC with Grant No
2007CB925204, and the Scientific Research Fund of Hunan Provincial
Education Department of China (Grant No. 07C579). We acknowledge the
useful discussions with Prof. C. P. Sun.

\appendix

\section{Factorization of unitary operator}

The shifted distance along the transverse direction in Eq.(\ref{4-13}) is
obtained by applying the unitary operator $U\left( t\right) $ to the initial
wavefunction of the probe field, where the unitary operator $U\left(
t\right) $ is generated by Hamiltonian $H$ in Eq. (\ref{4-12}). As operator $%
p_{z}$ commutes with other terms in Eq. (\ref{4-12}), the unitary operator $%
U\left( t\right) $ can be firstly factorized as
\end{subequations}
\begin{equation}
U\left( t\right) =e^{-iv_{g}tP_{z}}U^{\prime }\left( t\right),
\label{SG2ap-01}
\end{equation}%
where%
\begin{equation}
U^{\prime }\left( t\right) =\exp \left( -iH_{x}t\right)  \label{SG2ap-02}
\end{equation}%
is generated by%
\begin{equation}
H_{x}=\frac{1}{2m_{p}}p_{x}^{2}+\eta _{0}+\eta _{1}\left( x-a\right) \text{.}
\label{SG2ap-03}
\end{equation}%
It means that the unitary operator $U^{\prime }\left( t\right) $ only
contains operator $P_{x}$ and $x$, which generate the Lie algebra with the
basis $\{x,P_{x},P_{x}^{2},\mathbf{1}\}.$ Thus, operator $U_{li}^{\prime }$
can be factorized as the form
\begin{equation}
U^{\prime }\left( t\right)
=e^{g_{1}P_{x}^{2}}e^{g_{2}P_{x}}e^{g_{3}x^{\prime }}e^{g_{4}},
\label{SG2ap-04}
\end{equation}%
and $g_{i}=g_{i}(t)$ are unknown functions of time $t$ to be determined.
Here $x^{\prime }=x-a$.

Mathematically, the above factorization Ansatz is based on the Wei-Norman
algebraic theorem \cite{swna}: if the Hamiltonian of a quantum system%
\begin{equation}
H=\sum_{j=1}^{K}C_{j}(t)X_{j}  \label{SG2ap-05}
\end{equation}%
is a linear combination of the operators $X_{j}$ that can generate a $N$%
-dimensional Lie algebra with the basis:
\begin{equation}
\{X_{1},X_{2},...,X_{k},X_{k-1},....X_{N}\},  \label{SG2ap-06}
\end{equation}%
then the evolution operator governed by $H$ can be factorized as a product
of the single parameter subgroups, that is ,
\begin{equation}
U=\prod\limits_{j-1}^{N}e^{\xi _{j}(t)X_{j}},  \label{SG2ap-07}
\end{equation}%
where the coefficients $\xi _{j}(t)$ can be determined by the ``external
field parameters'' $C_{j}(t)$ through a system of non-linear equations.

Now, we differentiate (\ref{SG2ap-04}) with respect to $t$ and multiply the
resulting expression on the right hand side by the inverse of (\ref{SG2ap-04}%
), obtaining
\begin{subequations}
\label{SG2ap-08}
\begin{align}
& \frac{1}{2m_{p}}p_{x}^{2}+\eta _{0}+\eta _{1}x^{\prime }= \\
& i\frac{\partial g_{1}}{\partial t}P_{x}^{2}+iP_{x}\left( \frac{\partial
g_{2}}{\partial t}-2ig_{1}\frac{\partial g_{3}}{\partial t}\right)  \notag \\
& +ix\frac{\partial g_{3}}{\partial t}+i\left( \frac{\partial g_{4}}{%
\partial t}-ig_{2}\frac{\partial g_{3}}{\partial t}\right) .  \notag
\end{align}%
This leads to a systems of coupled differential equations
\end{subequations}
\begin{align}
& i\frac{\partial g_{1}}{\partial t}=\frac{1}{2m}\text{, }  \label{6-12} \\
& i\frac{\partial g_{3}}{\partial t}=-\eta _{1}\text{, } \\
& i\left( \frac{\partial g_{2}}{\partial t}-2g_{1}i\frac{\partial g_{3}}{%
\partial t}\right) =0, \\
& i\left( \frac{\partial g_{4}}{\partial t}-i\frac{\partial g_{3}}{\partial t%
}g_{2}\right) =-\eta _{0}.
\end{align}%
The solution to these equations reads
\begin{subequations}
\label{SG2ap-09}
\begin{align}
g_{1}& =-i\frac{t}{2m}\text{, } \\
g_{3}& =-i\eta _{1}t, \\
g_{2}& =-i\frac{\eta _{1}}{2m}t^{2}\text{,} \\
g_{4}& =-i\left( \eta _{0}t+\frac{t^{3}}{3}\frac{\eta _{1}^{2}}{2m}\right) .
\end{align}%
By applying the unitary operator $U\left( t\right) $ to the Gaussian probe
wave packet in Eq. (\ref{4-09}), a straightforward calculation shows that
the center of the probe wave packet is shifted to the position given in Eqs.
(\ref{4-13}).

\end{subequations}

\end{document}